\documentclass[aps, pre, floatfix,
reprint,groupedaddress,
showpacs
]{revtex4-1}

\usepackage{graphicx}
\usepackage{bm}

\usepackage{amsmath}
\usepackage{amssymb}
\usepackage{color}
\usepackage{subfigure}
\usepackage{times}
\usepackage{float}
\usepackage[sort&compress]{natbib}

\usepackage[colorinlistoftodos,prependcaption,textsize=small]{todonotes}

\usepackage{hyperref}

\hypersetup{
  colorlinks=true,
citecolor=black,
linkcolor=black,
urlcolor=black
  }

\newcommand{\wmax}{W_{\mathrm{max}}}
\newcommand{\smax}{S_{\mathrm{max}}}

\begin{document}

\title{Weighted Axelrod model: different but similar}

\author{Zuzanna Kalinowska}
\email{z.kalinowska@alumni.uj.edu.pl}

\affiliation{
Faculty of Physics, Astronomy and Applied Computer Science, Jagiellonian University, ul. St. {\L}ojasiewicza 11,
30--348 Krak\'ow, Poland
}

\author{Bart{\l}omiej Dybiec}
\email{bartlomiej.dybiec@uj.edu.pl}
\affiliation{
Institute of Theoretical Physics, and Mark Kac Center for Complex Systems
Research, Jagiellonian University, ul. St. {\L}ojasiewicza 11,
30--348 Krak\'ow, Poland
}

\begin{abstract}
The Axelrod model is a cellular automaton which can be used to describe the emergence and development of cultural domains, where culture is represented by a fixed number of cultural features taking a discrete set of possible values (traits).
The Axelrod model is based on two sociological phenomena: homophily (a tendency for individuals to form bonds with people similar to themselves) and social influence (the way how individuals change their behavior due to social pressure).
However, the Axelrod model does not take into account the fact that cultural attributes may have different significance for a given individual.
This is a limitation in the context of how the model reflects mechanisms driving the evolution of real societies.
The study aims to modify the Axelrod model by giving individual features different weights that have a decisive impact on the possibility of aligning cultural traits between (interacting) individuals.
The comparison of the results obtained for the classic Axelrod model and its modified version
shows that introduced weights have a significant impact on the course of the system development, in particular, increasing the final polarization of the system and increasing the time needed to reach the final state.
\end{abstract}


\maketitle


%
%
%
\section{Introduction and motivation\label{sec:introduction}}

Agent-based modeling \cite{bankes2002agent,jackson2017agent} and cellular automata \cite{wolfram2002} are two classical techniques which are typically used for the simulation of complex systems consisting of autonomous agents.
These methods have been significantly developed due to the increase in computational power and its availability.
They are applied in versatility of systems which are built by a big number of similar units (agents or players) ranging from epidemiological models \cite{tracy2018agent} to social systems \cite{bruch2015agent,macy2002factors} with extensions to hydrodynamics \cite{rothman2004lattice} and others.
Agent-based modeling is very popular in computational sociology \cite{jackson2017agent,weidlich2000} as it allows for examination of connections between microscopic dynamics and macroscopic properties.
It allows for verification of various hypotheses, which are assessed using in-silico experiments.
In particular it can be used for exploration of connections between microscopic dynamics, which can be hard to quantify, and the macroscopic, observable, properties (output) of social systems.
Therefore, it can be used to examination of emergence of collective behavior like segregation \cite{schelling1971dynamic}, flocking \cite{reynolds1987flocks,vicsek1995novel}, opinion formation \cite{hegselmann2002,galam2002,jedrzejewski2019statistical}, group formation \cite{castellano2009,gray2014emergence,schweitzer2021social},  development of cultural centers  \cite{dybiec2012b} and exploration of efficiency of game strategies \cite{axelrod1997evolution,press2012iterated}.
Cellular automata allow for verification of robustness of opinion-based groups \cite{dybiec2012c} and description of hierarchy maintenance \cite{czaplicka2014information,dybiec2014}.

The Axelrod \cite{axelrod1997,axelrod2021preventing} model is a well-known model of culture development and dissemination describing a possible mechanism for the emergence of cultural domains.
It is based on two sociological phenomena: homophily (tendency for individuals to form bonds with people similar to themselves) and social influence (the way how individuals change their behavior in response to social pressure) \cite{holley1975,lewenstein1992}.
Technically, it assumes that every culture is represented by a vector of $F$ cultural traits (features), each taking any of the $q$ allowed opinions (values).
The model assumes that an individual can interact with local neighbors if and only if they share some common traits.
The agents are conservative in the sense that they are more likely to interact with other agents who are similar to them.
This simplistic minimalist approach makes it already possible to observe a variety of social behaviors which can be encountered in real-life situations.

 On the one hand, at every successful interaction, one of the interacting agents accepts the agent’s point of view on a topic on which both agents differ.
 Consequently, interactions increase the similarity between agents and make them even more likely to interact in the future. On the other hand, acceptance of opinion can result in differentiation from other neighbors.
In the final state, the Axelrod model allows for coexistence of multiple cultural domains where neighboring cultures are completely different, as agents belonging to adjacent clusters do not share any common traits.

The original Axelrod model \cite{axelrod1997} assumes that initial similarity is the starting condition for any interaction.
It makes the same opinion on a simple, not controversial point, e.g., favorite color, which can result in alignment on a difficult topic, e.g., politics.
Such an extreme but possible case might occur because each cultural feature is equally important.
Moreover, with the increasing number of cultural features, the condition for interaction favors system homogenization, because for the increasing number of topics it is easier to find a point on which both agents agree.
Therefore, the common opinion on a given particular subject can result in full alignment of opinion.
This seems to be in contradiction with real-life observations where, despite imitation of neighbors' behavior, multiculturalism is preserved.
Consequently, multiple generalizations, increasing polarization, to the Axelrod model have been suggested.
For instance, they include a layered version of the Axelrod model \cite{battiston2017layered}, which reflects the fact that an individual belongs to more than one network and various issues/topics are discussed within different groups of neighbors.
Analogously, to start an interaction the full agreement on a given topic is not always required. It can be sufficient that states are close enough \cite{maccarron2020agreement} like in bounded-confidence models \cite{hegselmann2002,weisbuch2004bounded}.

The Axelrod model does not take into account the fact that cultural attributes may have different significance for a given individual.
This is a limitation in the context of how the model reflects mechanisms driving the evolution of real societies.
The current study aims to modify the Axelrod model by giving individual features different weights that have a decisive impact on the possibility of changing the opinion and in turn on interactions between two individuals.

%
%
%
\section{Model and results \label{sec:model}}

We start with the discussion of the classical Axelrod model \cite{axelrod1997}, afterwards we present the modified version.
The suggested extension to the Axelrod model assumes that every cultural feature is associated with a weight.
On the one hand, the weight defines the importance of the given cultural feature.
On the other hand, it defines the minimal level of similarity between neighbors which is required to adjust this particular cultural feature.
Such a constraint significantly changes overall model properties including its long-time behavior.

%
%
%
\subsection{Original Axelrod model}

Axelrod's model is a multi-agent cellular automaton.
The model consists of a set of agents that take on a finite collection of (discrete) states.
The state of an agent depends on the system's previous state and is determined through a set of rules.
Those rules describe the mechanics of the agent's interaction with other actors in the neighborhood.
Each agent $k$ is characterized by a vector $X(k)$ consisting of $F$ cultural attributes that can take any of the allowed $q$ values (traits):
\begin{equation}
  X(k)=(\sigma_1, \dots , \sigma_F),
\end{equation}
where $\sigma_i \in \{1, 2, \dots , q\}$ for~$i = 1, 2, \dots , F$.

The model dynamics can be described in the following steps:

\vspace{-8pt}
 	\begin{enumerate}
 	\item Choose a random agent $i$.
 	\item Choose a random neighbor $j$ (from the von Neumann neighborhood) of this agent.
 	\item Choose a random cultural attribute $f$ that will be the subject of interaction between the above agents. Note that the chosen attribute needs to have different values for both players ($\sigma_f(i) \neq \sigma_f(j)$).
 	\item Perform an interaction between the chosen agents based on their similarity -- agent $i$ takes over the value of the cultural attribute of its neighbor $j$
 	\begin{equation}
     	\sigma_f(i)=\sigma_f(j)
 	\end{equation}
 	with the probability $P_{ij}$ equal to the ratio of common attributes to all possible cultural traits $F$
 	\begin{equation}
     	P_{ij}=\frac{1}{F}\sum_{f=1}^N\delta_{\sigma_f(i), \sigma_f(j)},
     	\label{eq:pint}
 	\end{equation}
 	where $\delta_{\dots}$ is the Kronecker delta.
 	The probability $P_{ij}$ measures the similarity between agents $i$ and $j$.
 	It is equal to 0 if agents totally differ and it is equal to 1 if agents' cultures are exactly the same.
	\item Increase time by one.
 	\item Repeat steps 1 -- 5 (one repetition = one iteration) until one of the final conditions is met:
 	\begin{itemize}
  	\item Homogeneity -- all agents have the same values, i.e., their similarity is equal to 1 -- no more further interactions will introduce any changes to the system,
  	\item Polarization -- agents are split into subgroups (\textit{cultural clusters}) and the similarity between neighboring players is equal to 0 -- no further interactions are possible.
  	\item Time constraint -- simulation has exceeded the maximum number of allowed iterations.
 	\end{itemize}
	\end{enumerate}

To approximate a large-scale system -- virtual society -- the agents are placed on the 2D~lattice with periodic boundary conditions, i.e., borders are ``glued'' together.
In topological terms, the space made by two-dimensional periodic boundary conditions can be thought of as being mapped onto a torus.

%
%
%
\subsection{Weighted Axelrod model}

 In the modified Axelrod model, analogously like in the standard one, each agent $k$ is characterized by a vector $X(k)$ consisting of $F$ cultural attributes that can take any of the allowed $q$ values
\begin{equation}
 X(k)=(\sigma_1, \dots , \sigma_F),
\end{equation}
as well as a vector $W(k)$ representing the weights of respective attributes:
\begin{equation}
  W(k)=(w_1, w_2, \dots , w_F),
\end{equation}
where $\sigma_i \in \{1, 2, \dots , q\}$ and~$0 < w_i < 1$ for~$i = 1, 2, \dots , F$.
More precisely, we have assumed that weights follow a uniform distribution on $(0,\wmax)$, i.e., $w_f(i) \sim U (0,\wmax)$, where $\wmax$ is the additional parameter.
The assigned weights value the cultural attributes and make them more difficult to be changed.
They are also responsible for introduction of agents which are unlikely to alter a cultural feature characterized by a large weight.
These agents, are potential candidates for leaders \cite{lewenstein1992,kacperski2000phase} as they are more likely to convince others than to be fully changed.
However, they cannot be considered as fully inflexible zealots \cite{verma2014impact}, as part of their culture can be amended.

The only difference in the algorithm for the weighted version of the model is in the interaction step. In the original model, the interaction was solely driven by the similarity between two chosen actors.
In the modified version of the Axelrod model, step number 4 in the dynamics outlined in the above section can be defined as:
Perform an interaction between the chosen agents based on their similarity -- agent takes over the value of the cultural attribute of its neighbor with the probability equal to the ratio of common traits to all possible cultural attributes $F$ if the agent's weight for the chosen cultural attribute is smaller than the similarity, i.e.,
\begin{equation}
	w_f(i) < P_{ij},
	\label{eq:similarity}
\end{equation}
see Eq.~(\ref{eq:pint}).
The additional condition on the similarity, see Eq.~(\ref{eq:similarity}), reflects the fact that some cultural attributes are considered as more important and their value can be changed by a neighbor which is sufficiently convincing, i.e., appropriately enough similar.
For $\wmax<1/F$ the weighted Axelrod model is equivalent to the original one, while for $w>1-1/F$ it is not possible to align cultural features.
Therefore, for $\wmax \gg 1-1/F$, possibility of successful interactions can be significantly decreased.

%
%
%
\subsection{Results}

The weighted model was examined with a focus on the influence of the introduced modification on the temporal evolution of the system and the properties of final states.
The temporal evolution can be characterized by the time dependence of the order parameter $\eta$, the average number of cultures, and the average culture size.
Final states, for instance, can be characterized by the order parameter \cite{parisi1983order,castellano2009}, the probability distribution of number of cultures and the average time to reach the final state $\langle \mathcal{T} \rangle$.
As the order parameter, we use
\begin{equation}
	\eta = \frac{\langle \smax \rangle }{L^2},
\end{equation}
where $\smax$ is the size of the maximal cluster, i.e., the size of the largest (dominating) cultural domain, while $L^2 = L \times L$ is the total number of agents.
The order parameter measures the level of homogenization.
For $\eta=1$ the system is fully homogeneous, i.e., it is built by a single cluster, while for $\eta=0$ it is fully disordered.

\begin{figure}[!ht]
\centering
\begin{tabular}{c}
  \includegraphics[angle=0,width=0.75\columnwidth]{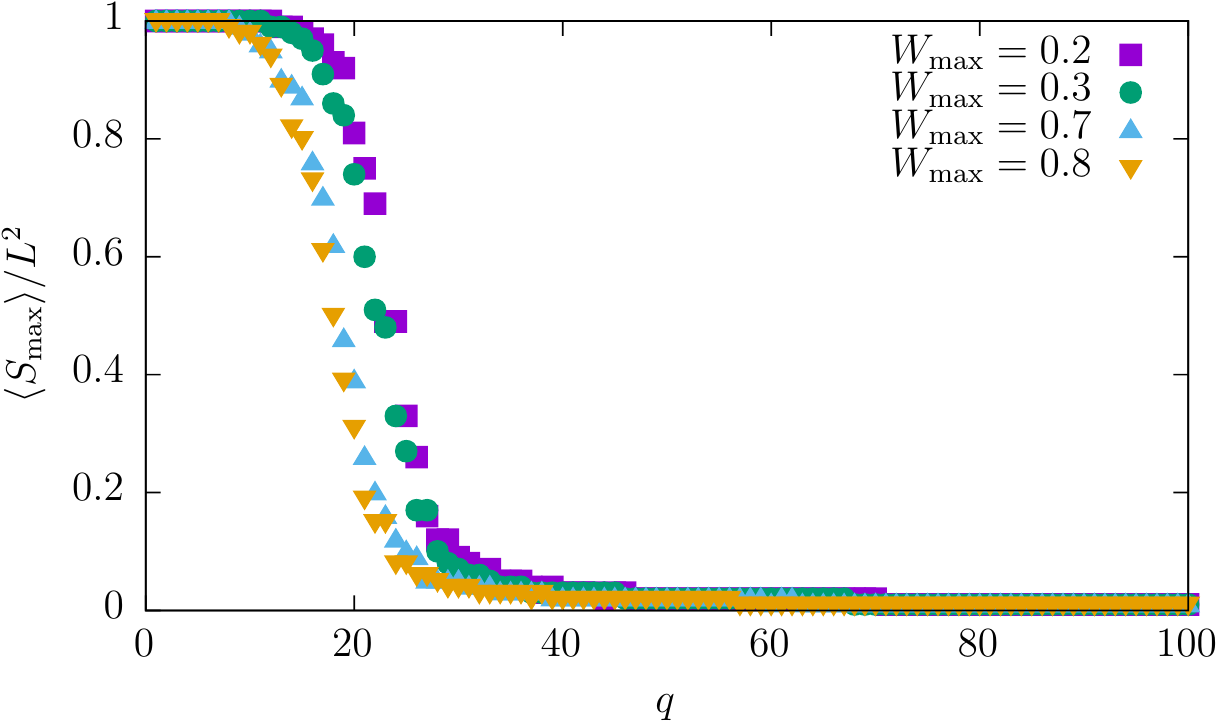} \\
\includegraphics[angle=0,width=0.75\columnwidth]{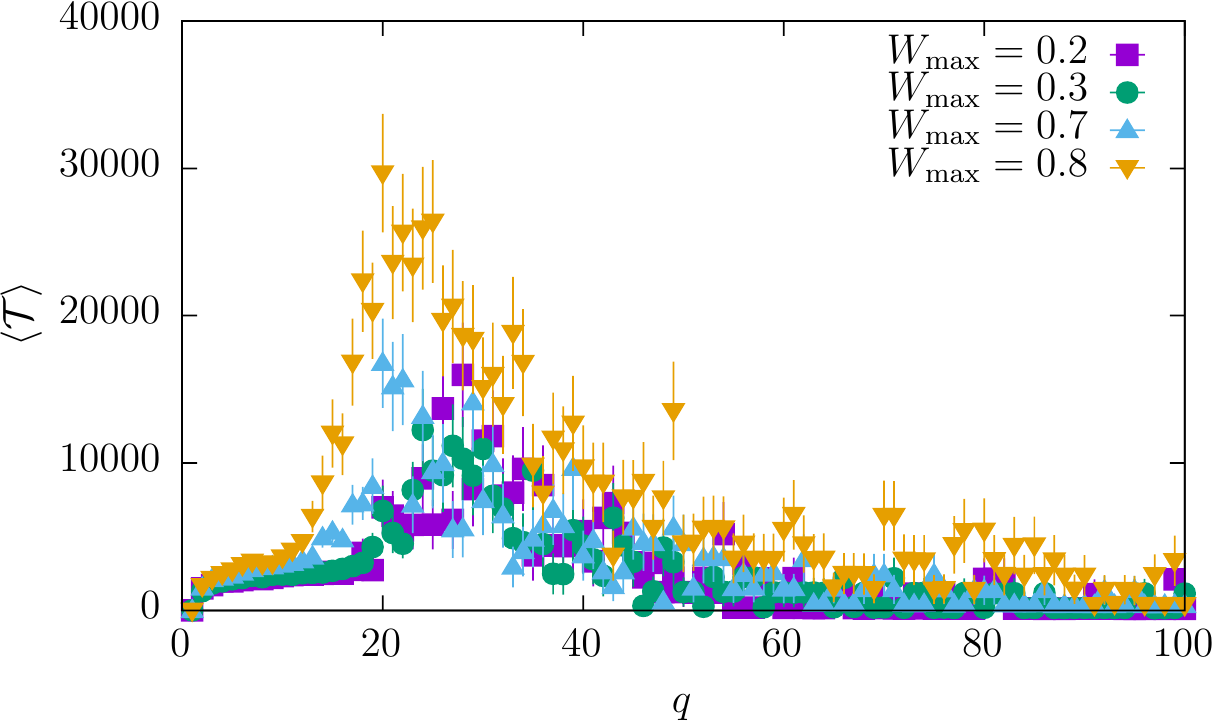} \\
\end{tabular}
  \caption{The order parameter $\langle \smax   \rangle / L^2$ (top panel) and the average time to reach the final state  $\langle \mathcal{T} \rangle$  as a function of $q$ for $L=20$ with $F=5$.
  Various curves correspond to different maximal weights $\wmax$, $\wmax\in\{0.2,0.3,0.7,0.8\}$.
  Results have been averaged over 100 realizations.}
  \label{fig:f5}
 \end{figure}

The standard Axelrod model is determined by two parameters: number of cultural features $F$ and the number of possible values  $q$.
It is generally known \cite{castellano2009} and intuitively understood that the increase in the number of cultural features $F$ increases chances of communication which in turn increases chances of system homogenization because for larger $F$ it is easier to find a common feature.
The increase in the number of possible values $q$ favors fragmentation, because for each point there are more possible options.
The studied extension to the Axelrod model introduces another parameter $\wmax$, a maximum value which the randomly chosen weight for the agent's cultural attribute could not exceed.
Within simulations, 400 agents were placed on the 2D $L \times L$ ($20 \times 20$) lattice with the periodic boundary conditions.
The von Neumann neighborhood was used in simulations, i.e., each agent interacts locally with its four nearest neighbors.
Each experiment was repeated 100 times so that it was possible to average results over multiple realizations. The maximum number of iterations was set to 100 000, which is large enough to assure that for the used system size the final state was indeed reached.
However, for the weighted Axelrod model, in addition to static final states, the final state can display some level of variability.
The possible variability originates due to competition between agents having at least one cultural attribute characterized by the high weight, which is unlikely to be changed.
These individuals can build cultural domains which are not separated by fully impenetrable boundaries, which are subject to fluctuations.

\begin{figure}[!ht]
\centering
\begin{tabular}{c}
  \includegraphics[angle=0,width=0.75\columnwidth]{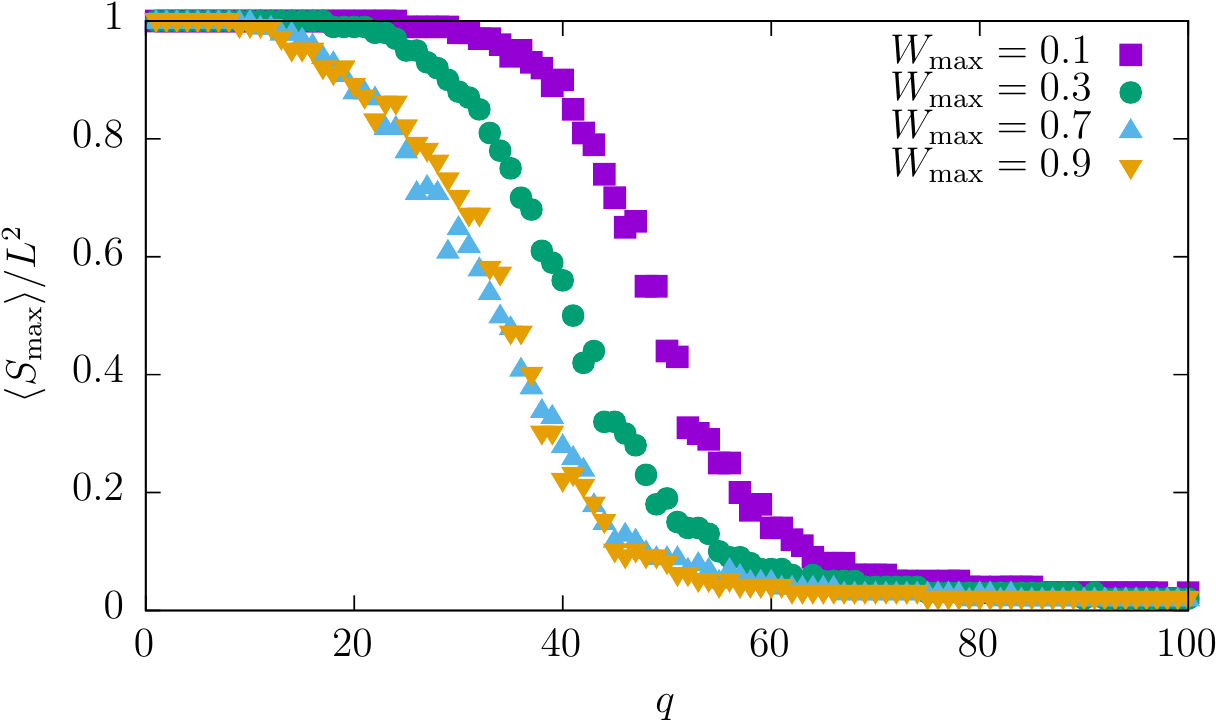} \\
\includegraphics[angle=0,width=0.75\columnwidth]{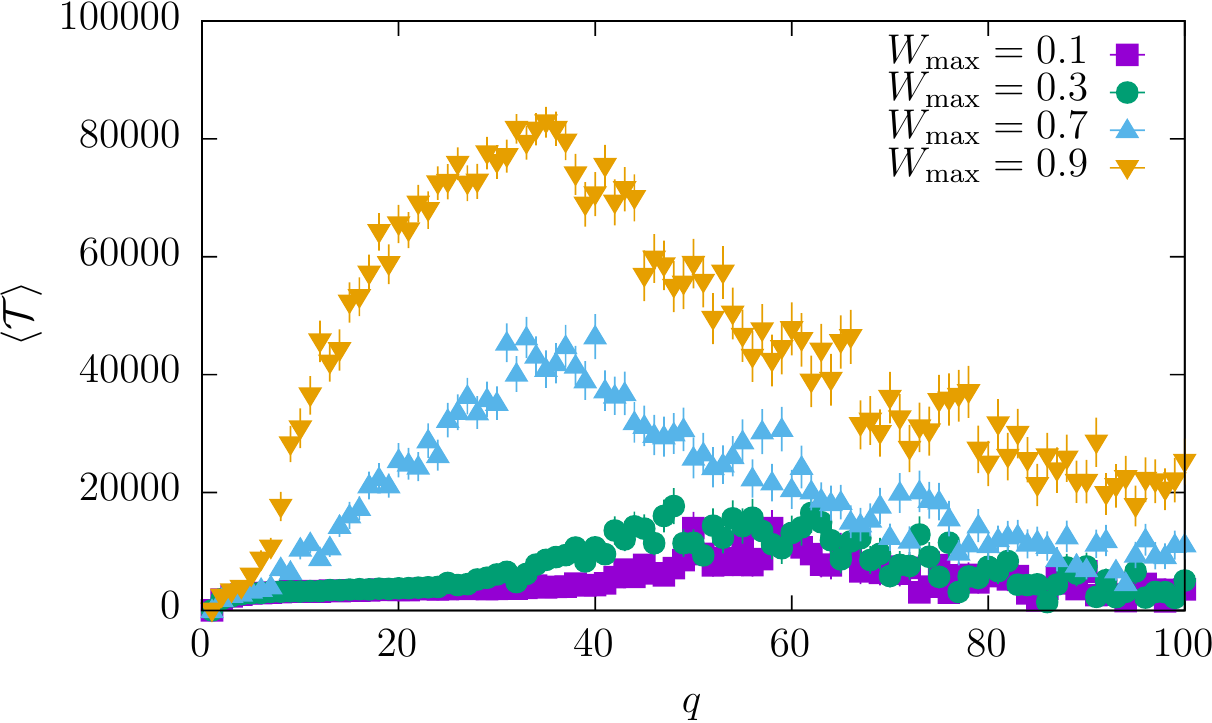} \\
\end{tabular}
  \caption{The same as in Fig.~\ref{fig:f5} for $F=10$ and $\wmax\in\{0.1,0.3,0.7,0.9\}$.}
  \label{fig:f10}
 \end{figure}

The modified model with $\wmax=1/F$ is equivalent to the original Axelrod model.
Generated weights, $w_f(i) \sim U(0,\wmax)$, are smaller than $1/F$ and the additional constraint imposed by weights is weaker than the requirement of having at least one common feature.
Therefore, simulations with $\wmax=1/F$ are (statistically) identical to the one for the original Axelrod model.
For $\wmax \gg 1-1/F$ individual weights $w_f(i)$ can be larger than $1-1/F$ and for these cultural features, the full adjustment of opinion is impossible.
To eliminate such a situation, we have assumed that $\wmax$ can be maximally equal to $1-1/F$, therefore, in principle, the possibility of perfect alignment is not eliminated.
Nevertheless, it is also possible to record a fragmented final state with coexisting cultural domains which could partially overlap.
The overlap is rooted in weights assigned to cultural attributes.
Neighboring clusters, built around two leaders characterized by large values of weights, can differ only with respect to features associated with large weights making clusters simultaneously different and similar.
Due to overlap the boundaries between such clusters are not static but relocate as agents on boundaries are likely to interact.
Furthermore, these interactions could result in emergence of ephemeral transient cultures positioned somewhere between cultures assigned to two competing clusters.

Introduction of weights increases the time needed to reach the final state because weights require an appropriate level of similarity and in turn slow down the adjustment of states as many interactions do not result in the adjustment of features.
In the weighted Axelrod model, the natural mechanism of culture dissemination is reconciliation by softening, i.e., first to adjust less controversial issues and then moving on to more problematic ones.
Reconciliation by softening is responsible for the growth of the time needed to reach the final state with the increase of $\wmax$.
Nevertheless, it still does not exclude complete homogenization of the system.
Interestingly, the problem of increasing time to reach the final states also has another less obvious side, i.e., the model could produce not fully static final states.
The assigned weights are not symmetric, i.e., for one of the neighboring agents a given cultural attribute can be very important, while for the second one, it can be a minor subject.
Therefore, the alignment of opinion on a given subject is not symmetric, as in one direction it can be easily performed while in the opposite direction, it can be difficult or even impossible to be carried out.
Large weights can be also responsible for the emergence of partially overlapping cultural domains.
However, these domains are subject to variation both in size and imputed culture because of asymmetry in interactions.
The size of overlapping adjacent domains constantly fluctuates because boundaries between them are not static.
The imputed culture can also evolve because some traits are easily modifiable because also a leader (agent with at least one large weight) can change some of cultural attributes and in turn the accepted culture.
In summary, for $\wmax>0$, the final state can be fully aligned or fragmented.
The fragmentation can be static (neighboring cultural domains do not overlap) or dynamic (neighboring cultural domains partially overlap).
In the dynamic case, robust fluctuations are visible.

\begin{figure}[!ht]
\centering
\begin{tabular}{c}
  \includegraphics[angle=0,width=0.75\columnwidth]{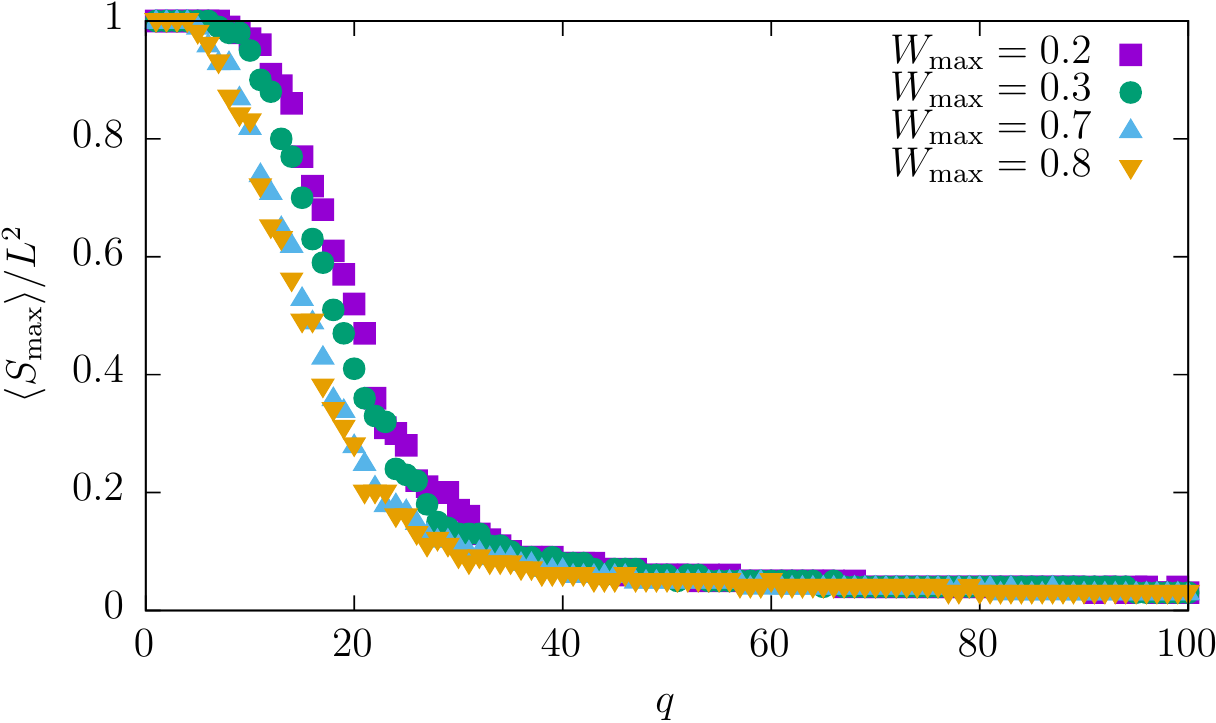} \\
\includegraphics[angle=0,width=0.75\columnwidth]{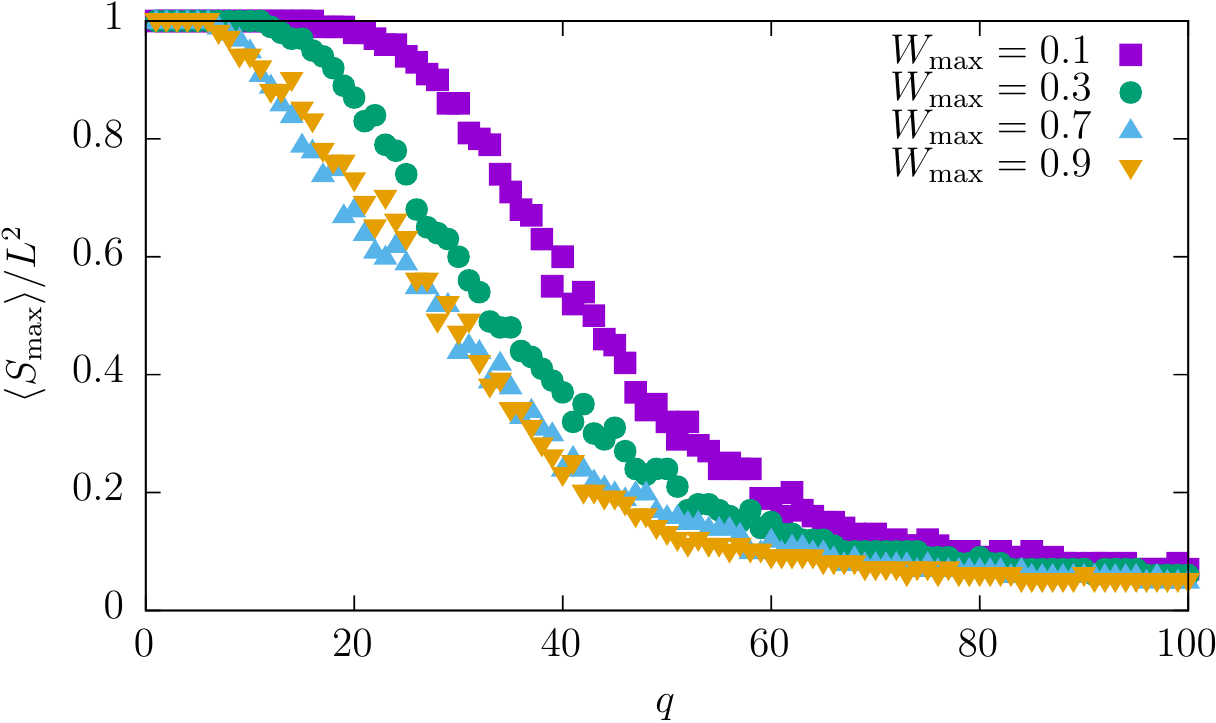} \\
\end{tabular}
  \caption{The order parameter $\langle \smax   \rangle / L^2$ for $F=5$ (top panel) and $F=10$ (bottom panel) with the smaller system size, i.e., $L=10$.
Various curves correspond to different maximal weights $\wmax$.
  Results have been averaged over 100 realizations.}
  \label{fig:l10}
 \end{figure}

 The top panel of Fig.~\ref{fig:f5} presents the dependence of the order parameter $\eta$ as a function of the number of possible values taken by individual features for $F=5$.
The $\langle \smax \rangle /L^2$ displays a standard pattern: for low $q$ the system is fully homogeneous ($\eta=1$).
 With increasing $q$ the system starts to fragment.
 For large $q$ the system is fully fragmented ($\eta \gtrapprox 0$).
 For all values of $\wmax$ the $ \eta (q)$ curves are similar, but the increase in $\wmax$ decrease the value of $q$ at which the order parameters starts to decay, i.e., for larger $\wmax$ the system is more likely to be in the disordered state.
 The bottom panel of Fig.~\ref{fig:f5} shows the average time $\langle \mathcal{T} \rangle$, measured in the number of iterations needed to reach the final state.
 For the fixed $q$ the average time $\langle \mathcal{T} \rangle$ is the growing function of maximal weight $\wmax$, because for large $\wmax$ larger similarity between agents to change their states is required.
 The increased required similarity needs a longer time to develop.

 \begin{figure}[!ht]
 \centering
\begin{tabular}{c}
  \includegraphics[width=0.75\columnwidth]{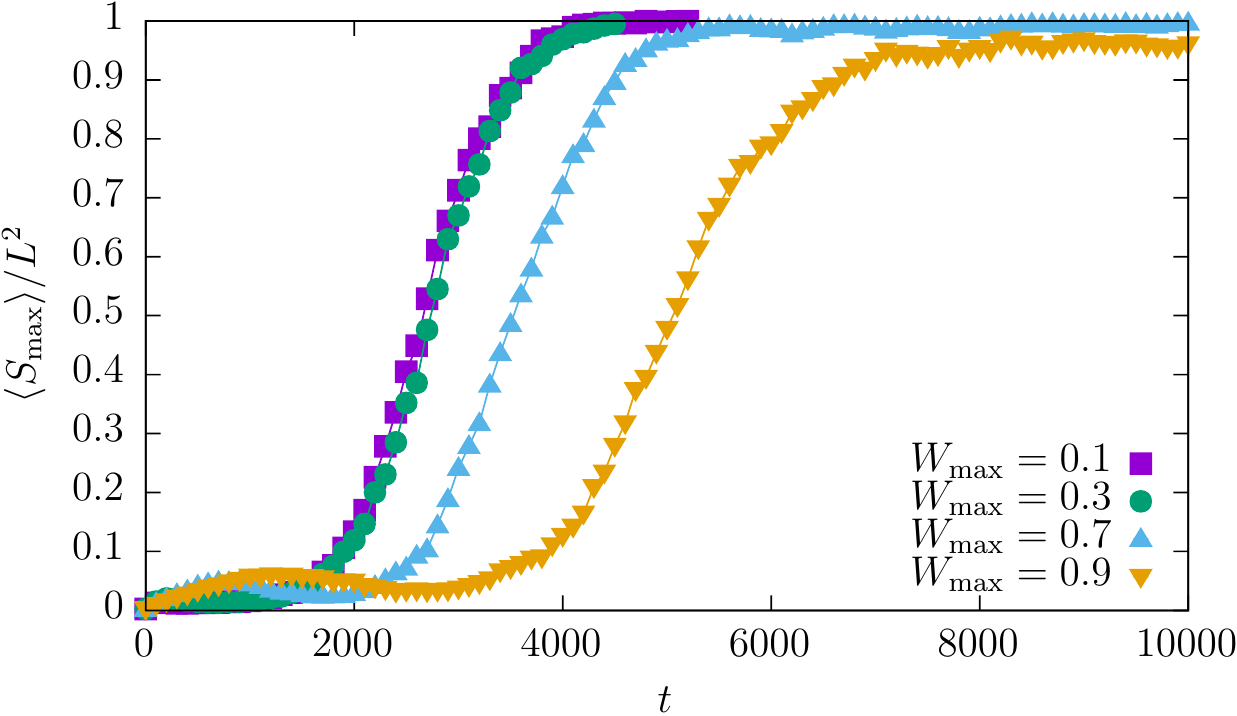}  \\
  \includegraphics[width=0.75\columnwidth]{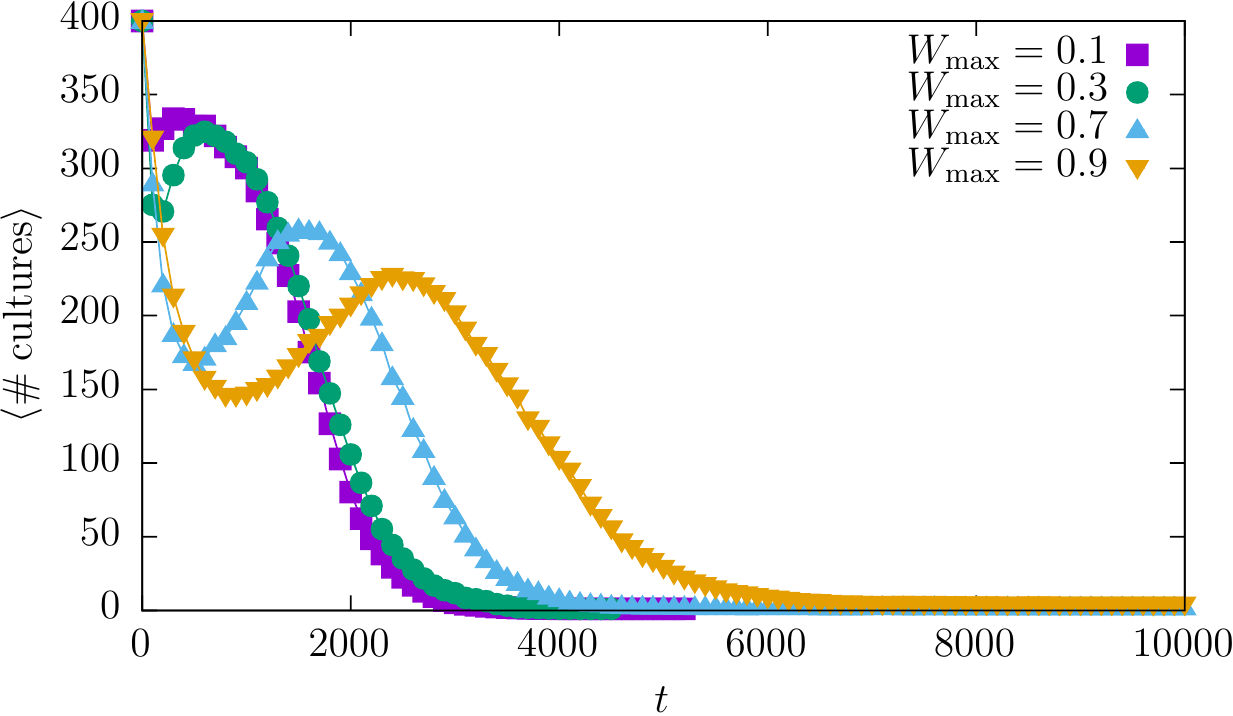}  \\
  \includegraphics[width=0.75\columnwidth]{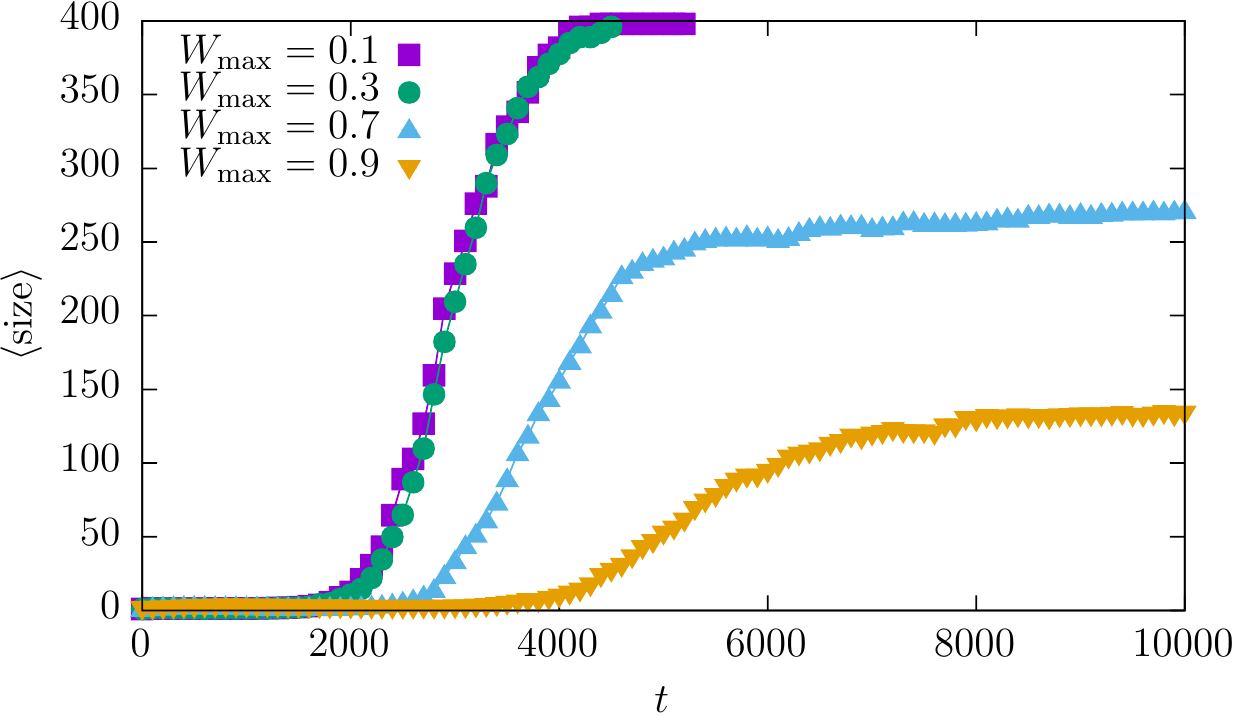}  \\
\end{tabular}
  \caption{Temporal characteristics of system dynamics for $400$ agents with $F=10$ and $q=10$.
  From top: the order parameter, i.e., the average maximal cluster size divided by the system size, the average number of cultures and the average culture size.
  Various curves correspond to different maximal weights $\wmax$, $\wmax\in\{0.1,0.3,0.7,0.9\}$.
  Results have been averaged over 100 realizations.
  }
  \label{fig:temporal-q10}
 \end{figure}

 \begin{figure}[!ht]
 \centering
\begin{tabular}{c}
  \includegraphics[width=0.75\columnwidth]{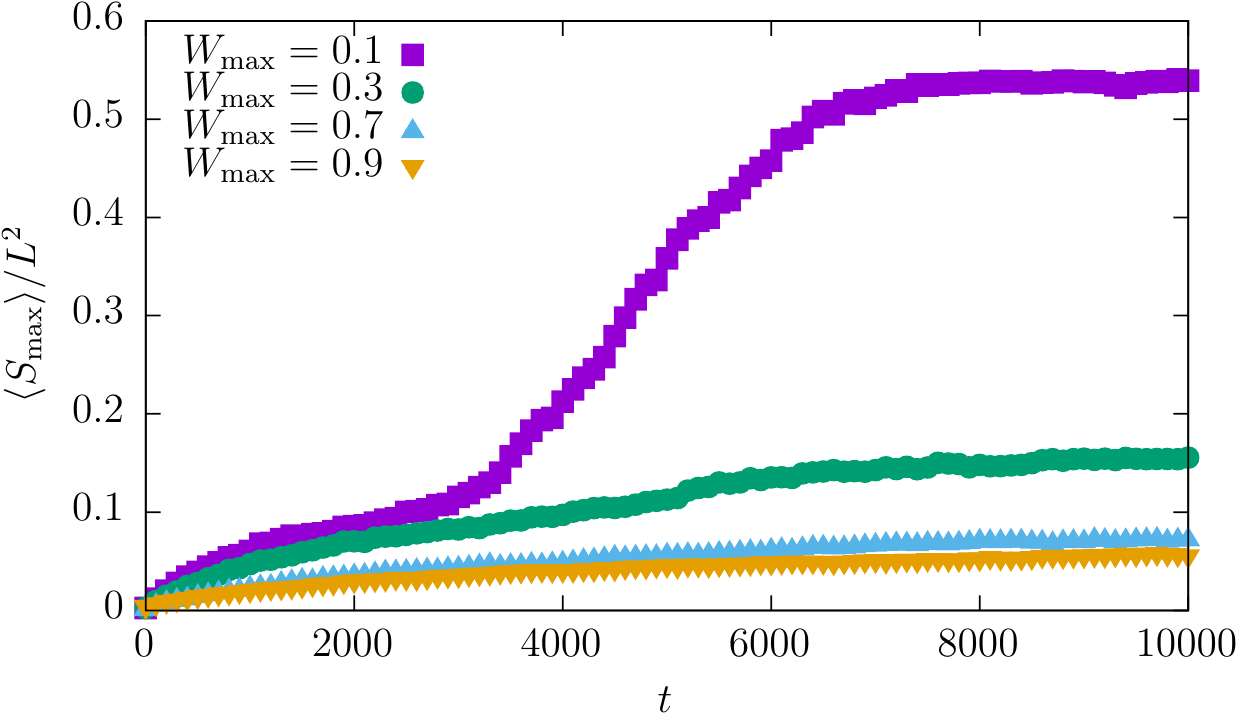}  \\
  \includegraphics[width=0.75\columnwidth]{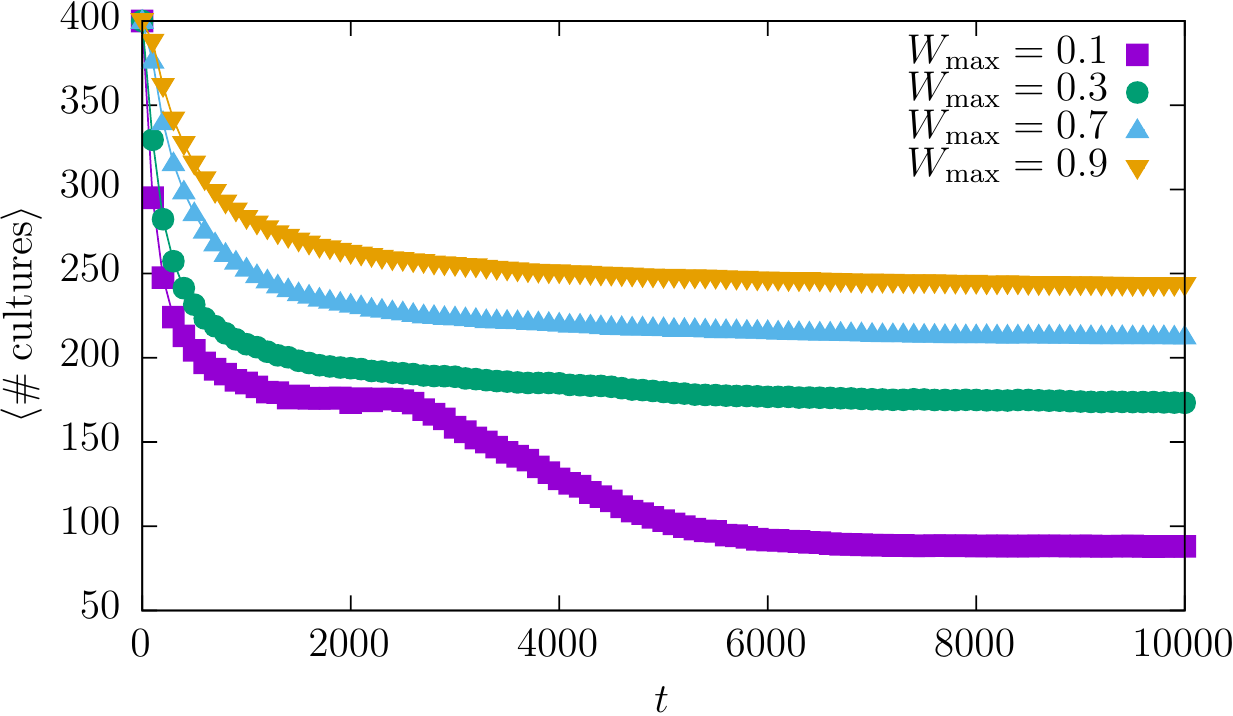}  \\
  \includegraphics[width=0.75\columnwidth]{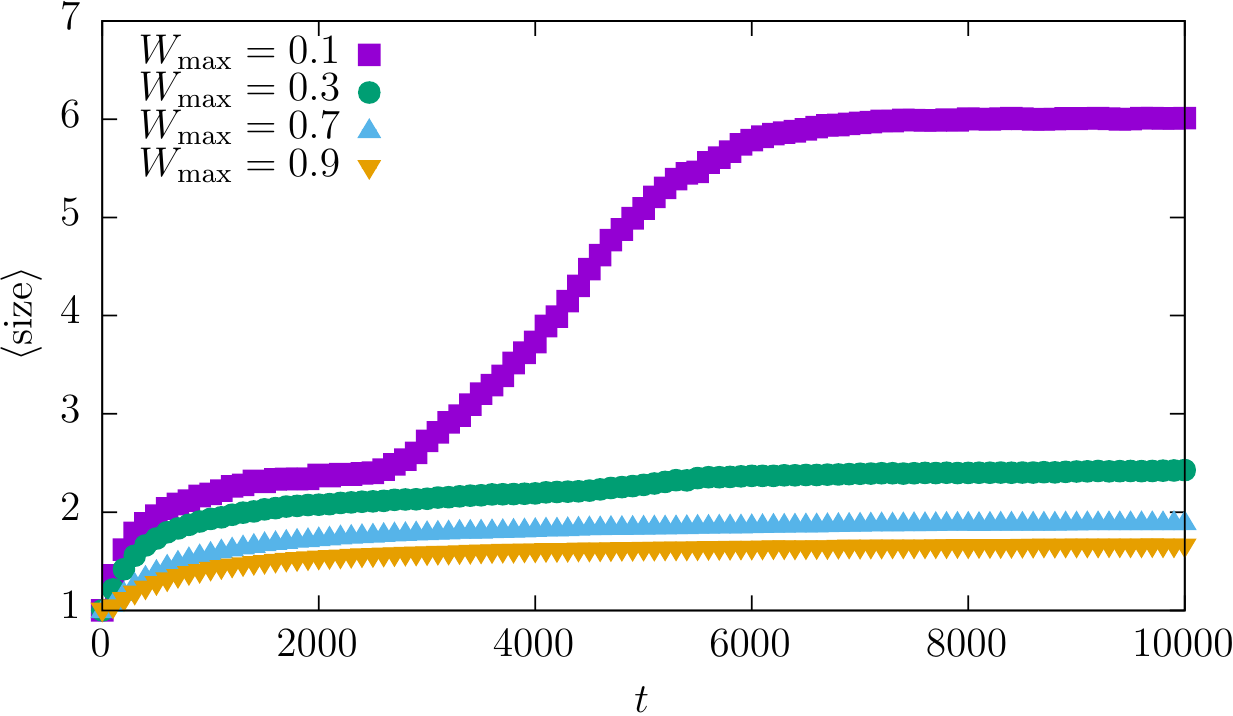}  \\
\end{tabular}
  \caption{The same as in Fig.~\ref{fig:temporal-q10} for $q=49$.}
  \label{fig:temporal-q49}
 \end{figure}

 Fig.~\ref{fig:f10}, in analogous way to Fig.~\ref{fig:f5}, displays the dependence of the order parameter and the average time needed to reach the final state for the two times larger number of cultural attributes, i.e., $F=10$.
 Due to increase in $F$, the domain where $\langle \smax \rangle /L^2 (q) \lessapprox 1$  has expanded because for larger $F$ it is easier to find common cultural traits.
 Also, differences between various $\wmax$ are more pronounced.
 Finally, the time needed to reach final states, in comparison to the $F=5$ case, has increased, as there are more features $F$ to be agreed on.

The majority of simulations have been performed for $400=20\times 20$ agents.
Such a choice gives a reasonable compromise between accuracy of results and simulation time.
Nevertheless, we have performed simulation for the smaller system, i.e., for the one consisting of $100=10 \times 10$ agents.
The order parameter, see Fig.~\ref{fig:l10}, is in accordance with results for the larger system size.
For a smaller system size, the transition between ordered and disordered state is more smooth.
More precisely, the range of $q$s where the order parameter drops from 1 to 0 has increased, which is the natural consequence of decreasing system size.

Examination of final states has been extended to exploration of the temporal behavior of vital system characteristics.
Figs.~\ref{fig:temporal-q10} and~\ref{fig:temporal-q49} present time dependence of the order parameter, the average number of cultures and the average culture size for $q=10$ (Fig.~\ref{fig:temporal-q10}) and $q=49$ (Fig.~\ref{fig:temporal-q49}).
$q=10$ lies in the domain where the final state is practically fully homogeneous, while for $q=49$ the final state is highly fragmented, see top panel of Fig.~\ref{fig:f10}.
For $q=10$, regardless of $\wmax$, the stationary state is practically fully aligned, but with the increasing $\wmax$ the time needed to reach the final state increases.
The number of cultures initially decays, however at some point, it starts to increase, assumes maximal value and drops down.
The local maximum of the number of cultures is the most pronounced for $\wmax=0.9$.
Such a non-monotonic dependence of $\langle \#\;\mathrm{cultures} \rangle$ can be attributed to local aligning of cultures among some agents which is interrupted by differentiation due to homogenization with other neighbors.
This effect is amplified by the increase in $\wmax$, because large weights are capable of stopping perfect local homogenization.
Adjacent partially overlapping cultures are vulnerable to changes on the interface and introduction of new transient cultures.
This effect is confirmed by the fact that number of cultures for $\wmax\in\{0.7,0.9\}$ has increased significantly but the order parameter has minimally decreased.
After some time the number of cultures starts to decay again as reconciliation by softening begins to work.
Finally, the average size, $\langle \mathrm{size} \rangle$, is a decreasing function of time.
However, the average size is very sensitive to the level of homogenization.
Since for $\wmax=0.7$ and $\wmax=0.9$ the order parameter is slightly smaller than unity, in addition to the dominating culture there are some very small clusters.
Therefore, the average size is significantly smaller than the system size.
For $q=49$ final states are disordered, see Fig.~\ref{fig:temporal-q49}, but the level of disorder depends monotonically on $\wmax$.
Average number of cultures, $\langle \#\; \mathrm{cultures} \rangle$, is a decaying function of time but it saturates at larger values than in the homogeneous case.
Due to system fragmentation the average culture size is small.

 \begin{figure}[!ht]
 \centering
\begin{tabular}{c}
  \includegraphics[width=0.75\columnwidth]{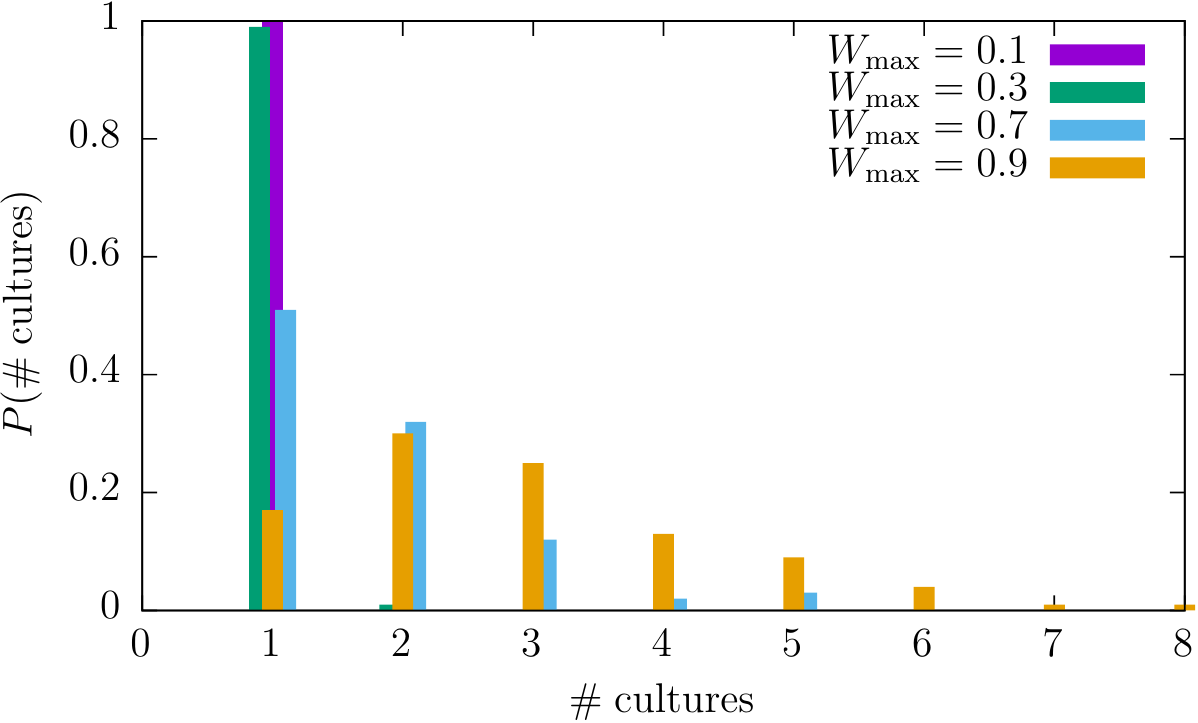}  \\  \includegraphics[width=0.75\columnwidth]{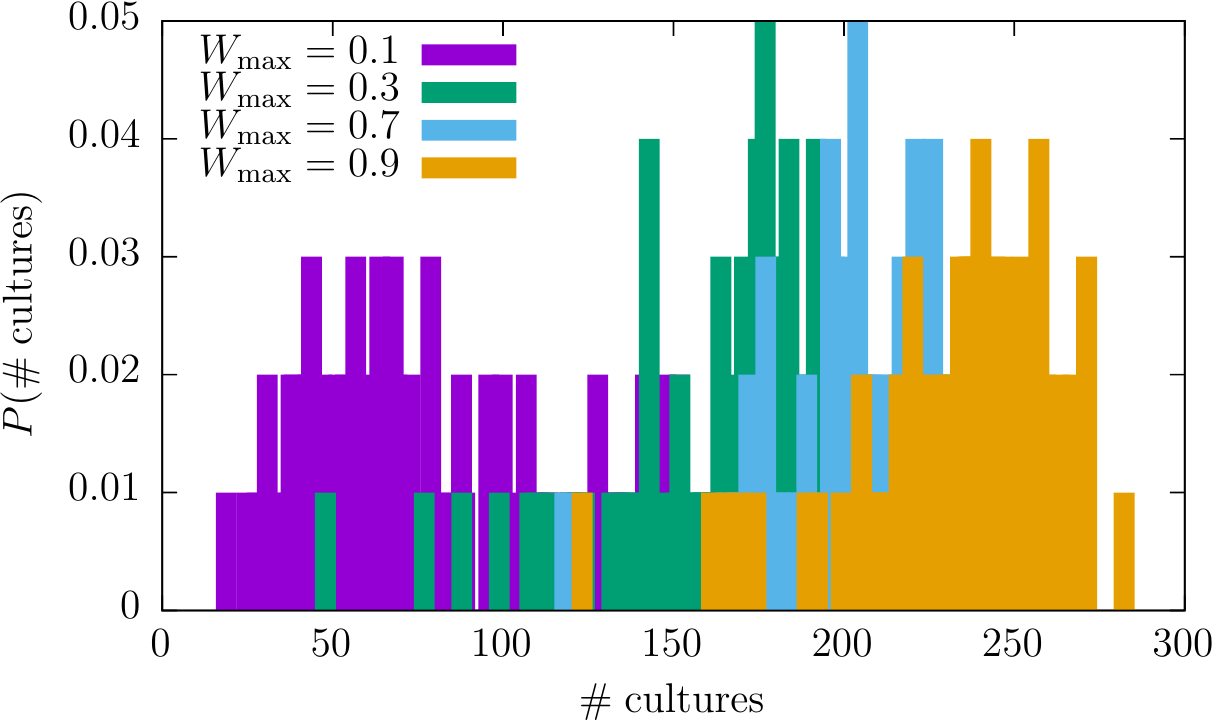}  \\
\end{tabular}
  \caption{The distribution of the number of cultures in the final state for $F=10$ with $q=10$ (top panel) and $q=49$ (bottom panel).
	Various curves correspond to different maximal weights $\wmax$, $\wmax\in\{0.1,0.3,0.7,0.9\}$.
  }
  \label{fig:pdf}
 \end{figure}

Finally, Fig.~\ref{fig:pdf} shows histograms of the number of cultures in the final state for $q=10$ (Fig.~\ref{fig:temporal-q10}) and $q=49$ (Fig.~\ref{fig:temporal-q49}).
From histograms, it is clearly visible that with the increasing $\wmax$ there is a larger dispersion of the number of cultures in the final state.
Moreover, as increasing $q$ decreases the order parameter, for larger $q$ there are statistically more cultures in the final states.
In the fragmented domain, the spread of the possible number of cultures can be very large, as the system is built out of many small domains.

%
%
%
\section{Summary and conclusions \label{sec:summary}}

The comparison of the results obtained for the classic Axelrod model and its modified version shows that the introduced weights have a significant impact on the system evolution, in particular, they typically increase the polarization of the system in the final state.
The final state is reached after a longer time and is built by a larger number of clusters.

The introduction of weights inevitably results in the increased system fragmentation, because similarity of agents can be not sufficient to align opinion of neighboring actors.
Therefore, in contrast to the original Axelrod model, cultural domains do not need to be separated by impenetrable boundaries, as large weights are capable of defining leaders which are ,,implementing'' their cultures.
Moreover, incorporation of weights increases the time needed to reach the final state, as not every level of similarity can produce further adjusting of cultures among neighbors.
The homogenization requires not only similarity but also a willingness to stand down, which is controlled by $\wmax$.
Readiness for the change decreases with growth in $\wmax$.
Consequently, the coexisting cultures can partially overlap, but thanks to persistence of their leaders they cannot homogenize.

Everyday observations suggest that neighboring groups, at the same time, could display significant similarity and maintain their diversity.
This feature cannot be reflected by the original Axelrod model, as neighboring cultures in the final state do not share any traits and, consequently, boundaries between cultural domains are impenetrable.
The weighted Axelrod model, with appropriately adjusted weights, is capable of capturing mentioned real-life situations.
Therefore, it can simultaneously promote some level of homogenization and maintain diversity.
Homogenization is reached on issues which are characterized by small weights, while diversity is maintained at points with large weights assigned.
The large weight-induced difference in some cultural attributes prevents full alignment of opinion and is responsible for emergence of cultural domains which are similar, but their followers (due to disagreement on a single issue) cannot unite.
Such partially overlapping domains are subject to variation in size as boundaries are subject to relocation.
Moreover, these groups can change the identifying motif, because a group leader (understood as a person strongly associated with one of cultural traits) is the one who defines the domain identity.
The leader can be persistent on some crucial (determined by large weights) issues, but can easily change their mind on others non-essential (determined by small weights) points.
Such a situation inevitably leads to emergence of similar but different cultural domains and ephemeral transient cultures.

%
%
\section*{Acknowledgements}

We gratefully acknowledge Poland’s high-performance computing infrastructure PLGrid (HPC Centers: ACK Cyfronet AGH) for providing computer facilities and support within computational grant no. PLG/2023/016175.
The research for this publication has been supported by a grant from the Priority Research Area DigiWorld under the Strategic Programme Excellence Initiative at Jagiellonian University.

%
%
%
\section*{Data availability}
The data (generated randomly using the model presented in the paper) that support the findings of this study are available from the corresponding author (ZK) upon reasonable request.

%
%
%

\def\url#1{}

\end{document}